\documentclass[twocolumn,10pt]{article}
\usepackage{epsfig,endnotes}
\usepackage{cleveref}
\usepackage{hyperref}
\usepackage[disable]{todonotes}
\usepackage{caption}
\usepackage{listings}
\usepackage{booktabs} 
\usepackage{subcaption} 
\usepackage{multirow}
\usepackage{amsmath}

\usepackage[tracking=true,kerning=true,spacing=true,final,activate=true]{microtype}
\SetTracking{encoding={*}, shape=sc}{0}

\usepackage[left=1.0in,right=1.0in,top=1.0in,bottom=1.0in]{geometry}		

\begin{document}

\date{}

\title{Don't Repeat Yourself:
Seamless Execution and Analysis of Extensive Network Experiments}

\author{
{Alexander Fr\"ommgen, Denny Stohr, Boris Koldehofe, Amr Rizk}\\
KOM - TU Darmstadt\\
\{firstname.lastname\}@kom.tu-darmstadt.de
}

\maketitle

\subsection*{Abstract}
This paper presents \emph{MACI},
the first bespoke framework for the management, the scalable execution, and the interactive analysis of a large number of network experiments.
Driven by the desire to avoid \emph{repetitive} implementation of \emph{just a few} scripts for the execution and analysis of experiments,
\emph{MACI} emerged as a \emph{generic} framework for network experiments that significantly increases efficiency and ensures reproducibility.
To this end, \emph{MACI}
incorporates and integrates established simulators and analysis tools to foster rapid but systematic network experiments.

We found \emph{MACI} indispensable in all phases of the research and development process of various communication systems,
such as \emph{i)} an extensive DASH video streaming study,
\emph{ii)} the systematic development and improvement of Multipath TCP schedulers,
and \emph{iii)} research on a distributed topology graph pattern matching algorithm.
With this work, we make \emph{MACI} publicly available to the research community to advance efficient and reproducible network experiments.

\section{Introduction}

Communication system research relies on experiments.
Accordingly, methods and tools, such as network simulators and their incorporated network models, emerged within the research community to enable controlled and repeatable experiments.
There are numerous simulators and emulators available.
These are tailored for different applications, underlying abstractions, and network models~\cite{afanasyev2012ndnsim, chan2014opennet, handigol2012reproducible, netravali2015mahimahi, osterlind2006cross, riley2010ns, 5999877, varga2008overview, zeng1998glomosim}.
Controlled experiments with these execution environments have become essential in the process of designing and developing communication systems to provide early and recurring feedback.

During our work on designing and developing different communication systems, we noted that we recurrently implemented support infrastructure and tools to automate experiments and analyze results.
The development of these tools typically started from scratch for every new research project. 
While we usually started with \emph{just a few scripts},
the tooling evolved with the research project,
and finally required a notable fraction of the overall research effort.
Although the development of such tools is straightforward,
it distracts from the actual research and delays the project.

In this paper, we identify three \emph{recurring} requirements for network experiment studies:
\emph{i)} the specification, management, and documentation of experiments with their dependent and independent control parameters,
\emph{ii)} the scalable experiment execution, i.e., the parallel execution of a large set of experiments,
and \emph{iii)} the interactive analysis of the experiment results based on the previously specified control parameters.
We argue that an \emph{integrated} solution is indispensable to increase the efficiency of network experiments.

In the following, we present \emph{MACI}, the first bespoke framework for the \emph{seamless} management, scalable execution,
and interactive analysis of a large number of experiments.
\emph{MACI} emerged as the result of our experiences and learned best practices during various research projects and evolved into a smart combination and integration of established tools to foster rigorous evaluations throughout the research process.
\emph{MACI} adopts, for example, the concepts of interactive data analysis from the domains of business intelligence and data science on network experiments.
\emph{MACI} follows the zeitgeist of agile development and continuous integration by removing obstacles to fast iterations which hinder research progress.

We discuss the benefits of \emph{MACI} based on our experience with three research projects:
\emph{i)} an extensive DASH video streaming study~\cite{multimedia},
\emph{ii)} the development of various Multipath TCP packet schedulers~\cite{progmp, thin_submitted},
and \emph{iii)} the tuning of a distributed topology graph pattern matching protocol~\cite{dtarl}.

We publicly release \emph{MACI} together with tutorials at \url{https://maci-research.net} to enable other researchers to increase the efficiency of their work.

\section{Requirement Analysis}
\vspace{-3mm}
To make the case for developing \emph{MACI},
we start by analyzing \emph{recent} observations and \emph{recurring} requirements for conducting network experiments.

\textbf{Req. 1: Improved Efficiency}
The driving requirement for an integrated network experiment framework is to improve research \emph{efficiency}.
This allows the researcher to focus on \emph{reasoning}, \emph{questioning} and \emph{improving} the observed behavior.

\textbf{Observation 1: Increasing Complexity}
While today's modular, layered communication systems enable optimizations and reduce complexity per layer,
research on communication systems has to consider complex cross-layer dependencies.
The tuning of transport protocols and congestion controls,
for example,
has to consider various network environments, application workloads, and configurations of the network stack.
Similarly, the performance of \emph{DASH} video streaming algorithms changes significantly when replacing the underlying TCP congestion control or transport protocol (e.g., replacing TCP with emerging protocols such as MPTCP and QUIC).
The systematic analysis of cross-layer dependencies is indispensable even if only a single component should be optimized.

\textbf{Observation 2: Increasing Innovation Speed}
We notice an increasing speed of network innovations.
The recently proposed QUIC transport protocol~\cite{quic}, for example,
is designed with the explicit goal of enabling frequent iterative improvements~\cite{langley2017quic}.
Hence, these \emph{iterative} improvements have to be \emph{repetitively} analyzed with respect to their impact on the application performance, e.g., in the previous \emph{DASH} video streaming example.
Recent advances in network programmability,
such as congestion control and Multipath TCP scheduler specification languages~\cite{arashloo2017hotcocoa, progmp},
will further increase innovation speed.
Since these languages enable rapid specifications of novel communication system algorithms,
we need support for rapid evaluations with systematic experiments.

\textbf{Observation 3: Extensive Experiments} We note an increasing number of extensive experiment studies in various communication system domains.
These extensive studies consist of a large number of individual emulation or simulation experiments.
Kakhki et~al.~\cite{longQuic} identify the need for rapid evaluations of protocols such as QUIC and
present a rigorous comparison of QUIC protocol versions.
Paasch et~al.~\cite{paasch2013benefits} used an experimental design approach for Multipath TCP to evaluate dependencies of the protocol configuration, the network capacity, and the network delay.
In~\cite{stohr2016qoe, Zabrovskiy:2017:AAV:3083187.3083221},
the authors conducted extensive emulation-based studies of DASH video streaming.
We found previously proposed experiment automation frameworks~\cite{Andreozzi:2009:FLS:1537614.1537718, Hallagan:2010:EAF:1808143.1808193, 5763591, 4654311, Perrone:2012:SSA:2429759.2430095} limited to network simulators such as ns-3.
Their deep integration makes these frameworks unsuitable various use cases, including the DASH and MPTCP studies in Section~\ref{sec:eval}.
All these examples confirm the need for extensive experiments and contribute frameworks for their confined research domain.
A \emph{general} \emph{reusable} experiment framework for communication systems research remains open.

\textbf{Observation 4: Resource Availability}
Evaluations with more experiment repetitions are usually favorable with regard to their insights and confidence but are time and resource consuming.
Recent infrastructure management advances pave the way for scalable experiment execution.
Tools such as OpenStack enable private clouds to easily allocate and share computing resources,
and public cloud providers have apparently infinite computing resources.

\textbf{Req. 2: Scalable, Parallel Experiment Execution}
The workload of network experiments with many configurations is embarrassingly parallel,
as there are no dependencies between the experiments~\cite{embarrassingly}.
Network experiment studies should leverage today's available experiment resources and the parallel nature of experiments to increase iteration speed.
The framework should reflect changing resource requirements during the research project lifecycle.

\textbf{Req. 3: Modular Framework}
The framework has to be modular to customize and exchange major components.
This includes APIs for additional components, e.g.,
to automatically trigger new evaluations based on previous results.
Network experiments require an execution environment such as a simulator, an emulator, a hardware testbed, or a real-world infrastructure.
Accordingly, it should be easy to integrate the variety of established execution environments.

\textbf{Req. 4: Interactive Analysis}
To foster a systematic analysis of the experiment results,
the framework has to manage the collection, aggregation, and analysis of results.
Following best practices from the areas of data analytics, business intelligence, and data science, data should be visualized interactively.
The researcher should \emph{interact} with the data to filter and aggregate for configurations and environments
and trigger the evaluation of additional configurations.

\textbf{Req. 5: Reproducibility}
The conducted scientific experiments must be reproducible.
This is particularly important as research prototypes evolve quickly and previous experiments have to be reproducible with their implementations and configurations.

\textbf{Req. 6: Coordination of Collaboration}
We notice that coordination of experiments and sharing of results among researchers introduces overhead.
Researcher tend to \emph{write just a small} analysis script,
as the development of reusable features is typically out of scope for the current research project.

\begin{figure*}[t]
\centering
\includegraphics[width=\linewidth]{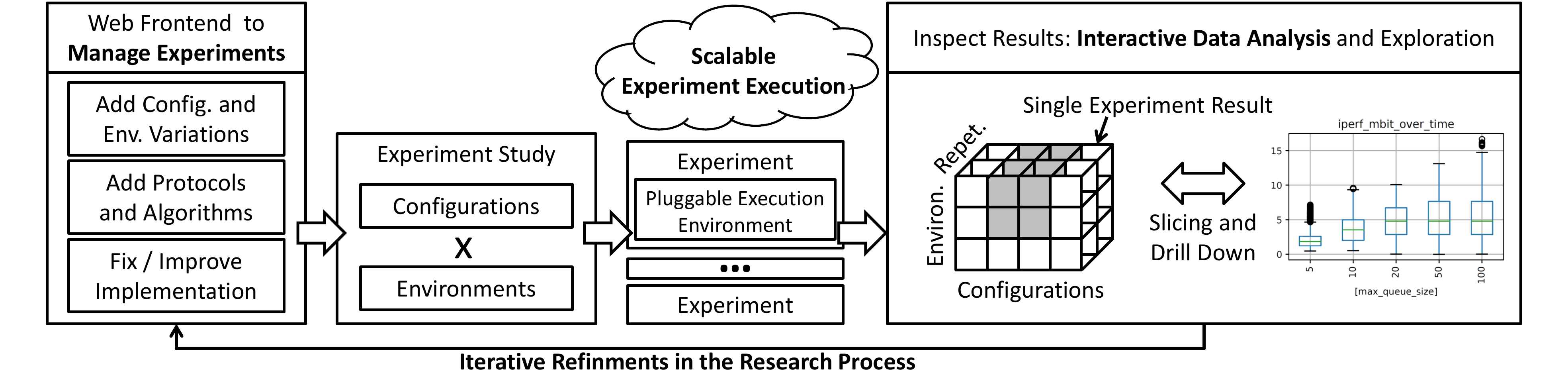}
\vspace{-6mm}
\caption{Overview of the experiment-driven research process enabled by \emph{MACI}.}
\vspace{-4mm}
\label{fig:maci_loop}
\end{figure*}

\section{Experiment-Driven Research}
\vspace{-2mm}

\emph{MACI} is designed for experiment-driven research,
which relies on recurring evaluations with implementations of systems, protocols, and algorithms.
In the following,
we present the design of \emph{MACI} for seamless experiment execution and interactive analysis.

\emph{MACI} supports the entire lifecycle of an iterative research process,
including the initial execution and analysis of prototypes with a few varying parameters,
the refinement of the underlying algorithms, protocols, and implementations,
and the extensive evaluation of matured implementations.
Therefore, \emph{MACI} enables the experiment management,
their scalable execution,
and the interactive analysis of the experiment results integrated in a seamless fashion, as shown in Fig.~\ref{fig:maci_loop}.

\begin{figure}[b]
\centering
\vspace{-4mm}
\includegraphics[width=\linewidth]{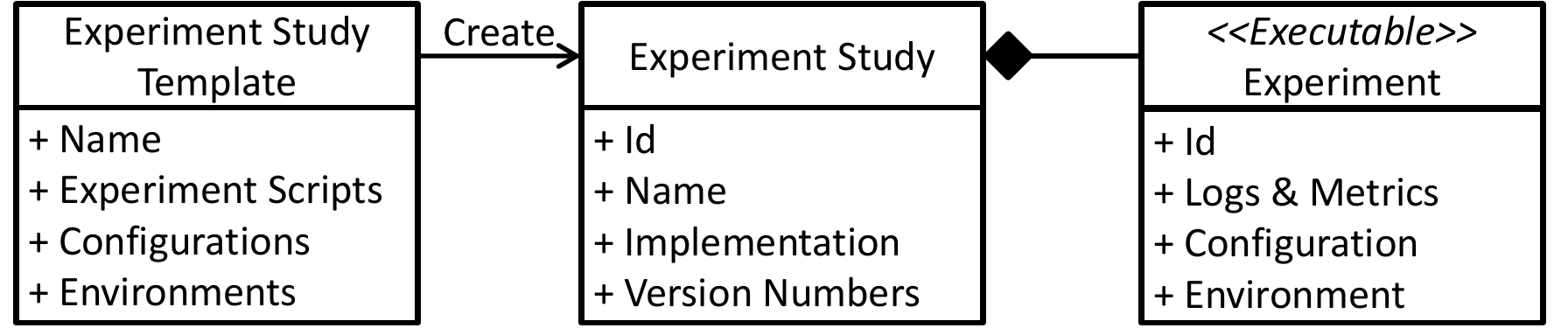}
\caption{An experiment study consists of multiple experiments with varying configurations.}
\label{fig:maci_model}
\end{figure}

\textbf{Manage Experiments}
\emph{MACI} structures experiments by decoupling \emph{experiment study templates}, \emph{experiment studies}, and \emph{experiments} to enable efficient management and reusability of experiments (Fig.~\ref{fig:maci_model}).
An experiment study template is a reusable template for a certain application domain.
The experiment study exposes dependency variables to control configuration and environment conditions.
Usually, evaluations compare the application performance in a certain environment depending on its configuration.
Accordingly, \emph{MACI} makes the differentiation between configuration and environment \emph{explicit} to automatically prepare for meaningful analysis.

An experiment study is a concrete instantiation of a template.
The experiment study comprises an \emph{executable} experiment, which results from the combinations of the specified configurations and environments.
The execution of a single experiment results in various measurements,
including target metrics and logging information.
The experiment script specifies the control flow and experiment process,
e.g., controlling tools such as ns-3, Mininet, or custom simulators.
\emph{MACI} keeps track of all meta information,
such as version number and commit identifiers of the used implementations,
to ensure reproducibility.

\textbf{Scalable Execution}
In \emph{MACI}, experiments are the smallest, atomic execution units.
\emph{MACI} controls the generation and parallel execution of \emph{experiments} in a scalable worker infrastructure.

\textbf{Interactive Data Analysis}
\emph{MACI} provides various views to interactively analyze experiment results.
These interactive views are \emph{seamlessly} available based on collected and provided data.
In particular,
the data model, e.g., the available configuration parameters, is automatically derived from the specified data in the management frontend.

The data analysis process is inspired by established features for the analysis of multidimensional data, i.e.,
the established OLAP (hyper) cube~\cite{codd1993providing, gray1997data}.
The user interface of \emph{MACI} allows the selection of target metrics, as well as the specification of filters and aggregations based on configuration and environment parameters.
The result of these operations is represented visually, e.g., as box plots.
The interactive analysis and visualization of the data distributions enables researchers to inspect sources of variances by changing filters and aggregations.

\emph{MACI} provides additional analysis views,
e.g., to analyze single experiments (\emph{drill down}) and balance conflicting target metrics.
The automatic generation of \emph{Pareto frontiers}, for example,
enables the researcher to inspect trade-offs for the throughput and latency of congestion controls.

\vspace{-3mm}
\section{Implementation}

In the following,
we present the modular implementation of \emph{MACI}.
The contribution of \emph{MACI} goes beyond these modules,
but stems from their seamless integration to foster the experiment-driven research process.

\textbf{Manage Experiments}
The web frontend includes an editor and management features for all steps of the experiment lifecycle,
i.e., the specification of the experiment and its configuration and environment parameters as well as the monitoring of running experiments.
The frontend provides direct feedback, e.g., the total experiment duration,
and automates reoccurring manual steps.
To integrate and control established network simulators and emulators, \emph{MACI} relies on Python scripts.
The backend is implemented as \emph{dotnet core} server application,
which provides a \emph{REST} API and a ready to use Java interface.

\textbf{Scalable Execution}
Experiment instances are executed in parallel to speed up the evaluation.
\emph{MACI} supports the manual management of worker instances (servers) as well as
the integration with manageable infrastructures, i.e., AWS EC2 and Proxmox.
The current implementation of \emph{MACI} follows an \emph{Infrastructure as a Service} cloud model,
as many experiments require own operating system modules (e.g., for transport protocol implementations such as MPTCP) and do not support multiple concurrent experiments per host.
For experiments with less infrastructure dependencies,
we envision more resource efficient serverless computations, such as AWS Lambda.

\textbf{Interactive Data Analysis}
For the data analysis,
we rely on the established \emph{SciPy}~\cite{scipy} data science tool-chain of \emph{Jupyter}, \emph{numpy}, and \emph{pandas}.
We discarded commercial alternatives in favor of a publicly available framework.
\emph{MACI} provides analysis template scripts which instantly provide interactive analysis features to explore and drill down experiments intuitively.
These templates are at the sweet spot of automation and flexibility,
as they are easily extendable by researchers with the vast Python software module ecosystem.

\textbf{Deployment}
To enable a rapid setup, we provide an optional docker-compose configuration,
initiating and connecting all required system components, i.e., the \emph{MACI-backend}, \emph{Jupyter/SciPy} and a \emph{Mininet worker}. 
Thus, a full \emph{MACI} system can be deployed with a single command on any major OS.

\vspace{-3mm}
\section{Experiences and Results}
\label{sec:eval}
\vspace{-2mm}

In the following,
we discuss our \emph{MACI} experiences.
We greatly benefited from \emph{MACI} during the development and evaluation in recent research projects on Multipath TCP scheduling~\cite{progmpdemo, progmp, thin_submitted, quic_submitted}, DASH video streaming~\cite{multimedia}, topology graph pattern matching algorithms~\cite{dtarl},
and the supervision of student theses.
We further reproduced the results of a notable Multipath TCP experimental design study~\cite{paasch2013benefits}.
Besides the necessary evaluation setup for the execution of a single experiment instance,
we only added six lines of code to benefit from all \emph{MACI} features,
such as the parallel experiment execution and the analysis with plots comparable to the original publication.

\textbf{Learning Curve}
We provided \emph{MACI} to students and found that \emph{MACI} \emph{i)} increased their speed and systematics by guiding them through the experiment lifecycle and \emph{ii)}  helped us to monitor their progress.

\textbf{Simulator/Emulator Integration}
While \emph{MACI} was developed with the Mininet network emulator in mind,
we integrated ns-3 and a custom Java-based simulator with minimal changes.

\vspace{-2mm}
\subsection{DASH Video Streaming Analysis}
\vspace{-1mm}

We used \emph{MACI} for an extensive \emph{Dynamic Adaptive Streaming over HTTP} (DASH) player and adaptation algorithm comparison.
While the results of this comparison are published in~\cite{multimedia},
we discuss the contribution of \emph{MACI} on this publication in the following.

\textbf{DASH}~\cite{DASH2011} is a main enabler of adaptive video streaming in today's Internet.
By adapting the quality and size of the downloaded video segments,
DASH copes with the wide range of fluctuating network conditions in today's Internet.
Various DASH players and video quality adaptation algorithms were proposed to provide high video playback quality and to avoid video stallings in these heterogeneous environments.

\begin{table}[t]
	\small
	\centering
	\caption{DASH study experimental design in~\cite{multimedia}.}
	\vspace{-3mm}
	\label{tab:emulation_vars}
	\begin{tabular}{c @{\hspace{1.5\tabcolsep}} c @{} c}
		\toprule 
		& {Variable} & {Values}  \vspace{-1mm} \\  \midrule
\multirow{4}{*}{Config.}		& Player & DASH.JS, Shaka, AStream\\
								& Adapt. Algo. & Standard, BOLA \\
								& Segment Length & 1, 2, 6, 10, 15 [s]\\
								& Target Buffer & Default, 5, 20 [s]\\ \midrule
                Env.      & $\mu_{\text{BW}}$ & 0.8, 2, 5, 7.5, 10 [Mbps]\\
							(BW)	&  $\sigma_{\text{BW}}^2$ & 0, 0.8, 2, 5 [$\text{Mbps}^2$]\\
	\bottomrule
	\end{tabular}
	\vspace{-2mm}
\end{table}

\textbf{Experiment Setup}
We used \emph{MACI} for a comprehensive DASH emulation study.
We compared three major DASH player implementations with two playback quality adaptation algorithms and various player configurations,
i.e., the video segment length and the target size of the playback buffer,
in networks with varying characteristics (Table \ref{tab:emulation_vars}).
For a detailed investigation,
we collected various target metrics,
including the achieved video quality,
the experienced stalling events,
and the network utilization.

\textbf{Iterative Research Process} We developed, tested, and improved the DASH specific measurement features iteratively.
The interactive analysis of the experiment results enabled us \emph{i)} to quickly detect errors and inconsistencies in our measurements and implementations and \emph{ii)} to identify regions of interested and to add additional measurement metrics and configurations to further investigate and question our findings within the process.
We profited from \emph{MACI} for interactive analysis group sessions to discuss and question hypotheses.
The simple repetition of experiment studies with improved and extended implementations was crucial for our efficiency.

\textbf{Scalable Execution}
As a single execution of all configurations in all environments requires more than 40 hours (120 s video playback per experiment),
the parallel experiment execution significantly increased our iteration speed and enabled us to retrieve reliable results with dozens of repetitions.

\vspace{-2mm}
\subsection{MPTCP Scheduler Development}
\vspace{-2mm}

We used \emph{MACI} for the development of five novel Multipath TCP (MPTCP) schedulers.
While the MPTCP specific details and evaluations are published in \cite{progmp, thin_submitted},
we discuss the contribution of \emph{MACI} on the design of one exemplary scheduler in the following.

\textbf{MPTCP}~\cite{rfc6824} is a recent TCP evolution,
which uses multiple \emph{subflows} to leverage multiple paths and network interfaces for a single connection.
The mapping of packets on subflows, the MPTCP scheduling,
has a crucial impact on the performance.
The design of MPTCP schedulers has to consider complex dependencies between subflow and traffic flow characteristics.

\begin{figure}[t]
\centering
\includegraphics[width=\linewidth]{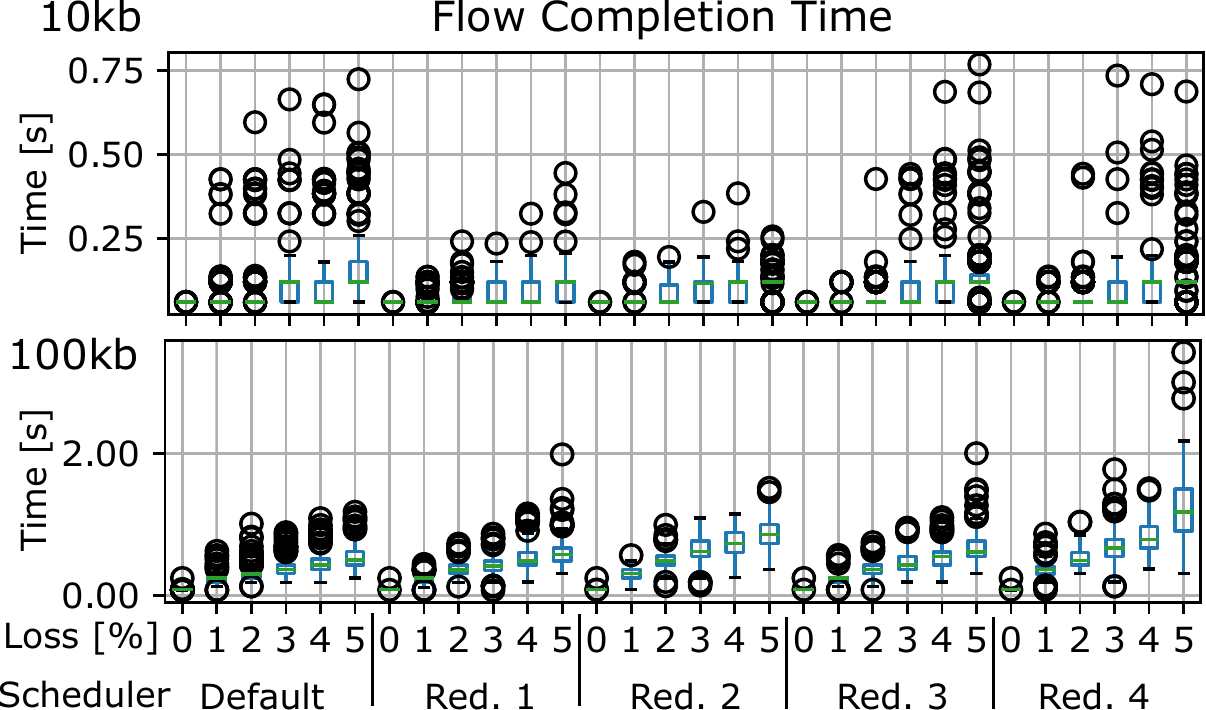}
\vspace{-6mm}
\caption{Excerpt of the systematic comparison of MPTCP scheduler redundancy flavors for different flow sizes (10kb, 100kb) and packet loss rates.}
\vspace{-4mm}
\label{fig:progmp}
\end{figure}

\textbf{Iterative Research Process} Redundant transmission of packets on multiple subflows proactively compensate packet loss and promises to reduce flow completion times.
Tuning a redundant scheduler,
however, calls for many design decisions, e.g., 
when to transmit a redundant or a fresh packet.
We used \emph{MACI} for a systematic comparison of these design decisions for various traffic patterns (e.g., flow sizes) in different network environments (e.g., loss rates and capacities).
The interactive analysis of \emph{MACI} with visualizations as shown in Fig.~\ref{fig:progmp} enabled us to identify and overcome weaknesses of scheduler designs.

\vspace{-4mm}
\section{Discussion}
\vspace{-2mm}

\textbf{I prefer simulator \emph{foo} and analysis tool \emph{bar}.}
\emph{MACI} focuses on a seamless experiment execution and evaluation process with established, publicly available components.
As there is no optimal tool for all scenarios,
the modular architecture of \emph{MACI} enables the integration of additional components, such as simulators and analysis tools.
For example, even though big data analysis frameworks were unrequired for our use cases so far,
\emph{MACI} supports their integration in the seamless research process.

\textbf{Isn't this just parameter sweeping?}
\emph{MACI} differs from parameter tuning and performance analysis frameworks~\cite{DuplyakinBrownRicci:Cluster16},
as it covers the entire research process, including the refinement of the evaluated protocols, algorithms, implementations, and their environments and configurations (Fig.~\ref{fig:maci_loop}).
\emph{MACI} increases the evaluation efficiency to focus on the analysis of research hypotheses and provide empirical evidence.

\textbf{Isn't this data dredging?}
The simplicity of conducting additional experiments and interactive analysis might be tempting to uncover statistically significant yet obviously unreasonable relations.
We claim, however, that researchers using \emph{MACI} save time to focus on rigorous analysis and work on better models.

\textbf{Isn't A/B testing superior?}
A/B tests~\cite{langley2017quic, schermann2016bifrost} are indubitably superior to emulation and simulation studies.
However,
rigorous and meaningful A/B testing \emph{i)} is reserved for a few leading companies and largely infeasible in academia and
\emph{ii)} requires systematic initial experiments which benefit from \emph{MACI}.

\vspace{-3mm}
\section{Conclusion}
\vspace{-2mm}

In this paper,
we presented \emph{MACI},
a framework for the management, the scalable execution, and the interactive analysis of a large number of network experiments.
\emph{MACI} significantly reduced repetitive tasks and increased the quality of the obtained results in various application scenarios~\cite{thin_submitted, progmpdemo, progmp, dtarl, multimedia, quic_submitted}. 
\emph{MACI} provided all evaluation process specific functionalities and allowed us to focus on research.
This paper provides only an overview of \emph{MACI}\textemdash{}many additional helpful features can be found in the released version.

\emph{MACI} is designed and evaluated with a focus on the experiences and requirements of researchers in the communication systems community.
We assume that the significance of \emph{MACI} and the idea of a seamless, integrated research process goes beyond this domain.
We released \emph{MACI} at \url{https://maci-research.net} and hope that it is the starting point to \emph{i)} increase the research efficiency and quality and \emph{ii)} integrate and establish more sophisticated evaluation methodologies in the communication system research process.

\vspace{-6mm}
\section*{Acknowledgment}
\vspace{-3mm}
\footnotesize
This work has been funded by the German Research Foundation (DFG) as part of the projects C2, C3, and B4 in the \emph{Collaborative Research Center (SFB) 1053 MAKI}.
This work was supported by the \emph{AWS Cloud Credits for Research} program.

\bibliographystyle{acm}
\bibliography{thesis}

\begin{thebibliography}{10}

\bibitem{scipy}
{Scripy 0.9.3: Python tools for manage system commands as replacement to bash
  script}.
\newblock Python Software Foundation \url{https://pypi.python.org/pypi/Scripy}.

\bibitem{afanasyev2012ndnsim}
{\sc Afanasyev, A., Moiseenko, I., Zhang, L., et~al.}
\newblock {ndnSIM: NDN simulator for NS-3}.
\newblock {\em University of California, Los Angeles, Tech. Rep\/} (2012).

\bibitem{Andreozzi:2009:FLS:1537614.1537718}
{\sc Andreozzi, M.~M., Stea, G., and Vallati, C.}
\newblock A framework for large-scale simulations and output result analysis
  with ns-2.
\newblock In {\em SIMUTools\/} (2009).

\bibitem{arashloo2017hotcocoa}
{\sc Arashloo, T., Ghobadi, M., Rexford, J., and Walker, D.}
\newblock {HotCocoa: Hardware Congestion Control Abstractions}.
\newblock In {\em HotNets\/} (2017).

\bibitem{chan2014opennet}
{\sc Chan, M.-C., Chen, C., Huang, J.-X., Kuo, T., Yen, L.-H., and Tseng,
  C.-C.}
\newblock {OpenNet: A simulator for software-defined wireless local area
  network}.
\newblock In {\em Wireless Communications and Networking Conference (WCNC)\/}
  (2014), IEEE, pp.~3332--3336.

\bibitem{codd1993providing}
{\sc Codd, E.~F., Codd, S.~B., and Salley, C.~T.}
\newblock {Providing OLAP (on-line analytical processing) to user-analysts: An
  IT mandate}, 1993.

\bibitem{DuplyakinBrownRicci:Cluster16}
{\sc Duplyakin, D., Brown, J., and Ricci, R.}
\newblock Active learning in performance analysis.
\newblock In {\em Proceedings of the IEEE Cluster Conference\/} (Sept. 2016).

\bibitem{rfc6824}
{\sc Ford, A., Raiciu, C., Handley, M., and Bonaventure, O.}
\newblock {{TCP Extensions for Multipath Operation with Multiple Addresses}}.
\newblock RFC 6824, 2013.

\bibitem{embarrassingly}
{\sc {Foster, Ian}}.
\newblock {\em {Designing and Building Parallel Programs}}.
\newblock {Addison–Wesley}, {1995}.

\bibitem{thin_submitted}
{\sc Fr\"{o}mmgen, A., Heuschkel, J., and Koldehofe, B.}
\newblock {Multipath TCP Scheduling for Thin Streams: Active Probing and
  One-way Delay-awarness}.
\newblock In {\em ICC\/} (2018).

\bibitem{progmpdemo}
{\sc Fr\"ommgen, A., and Koldehofe, B.}
\newblock {Demo: A Programming Model for Application-defined Multipath TCP
  Scheduling}.
\newblock In {\em ACM/IFIP/USNIX Middleware\/} (2017).

\bibitem{progmp}
{\sc Fr\"{o}mmgen, A., Rizk, A., Erbsh\"au{\ss}er, T., Weller, M., Koldehofe,
  B., Buchmann, A., and Steinmetz, R.}
\newblock {A Programming Model for Application-defined Multipath TCP
  Scheduling}.
\newblock In {\em ACM/IFIP/USNIX Middleware\/} (2017).

\bibitem{gray1997data}
{\sc Gray, J., Chaudhuri, S., Bosworth, A., Layman, A., Reichart, D.,
  Venkatrao, M., Pellow, F., and Pirahesh, H.}
\newblock {Data Cube: A relational aggregation operator generalizing group-by,
  cross-tab, and sub-totals}.
\newblock {\em Data mining and knowledge discovery\/} (1997), 29--53.

\bibitem{Hallagan:2010:EAF:1808143.1808193}
{\sc Hallagan, A., Ward, B., and Perrone, L.~F.}
\newblock {An Experiment Automation Framework for NS-3}.
\newblock In {\em SIMUTools\/} (2010).

\bibitem{quic}
{\sc Hamilton, R., Iyengar, J., Swett, I., and Wilk, A.}
\newblock {{QUIC}: A {UDP}-based secure and reliable transport for {HTTP/2}},
  July 2016.
\newblock IETF, Internet-Draft.

\bibitem{handigol2012reproducible}
{\sc Handigol, N., Heller, B., Jeyakumar, V., Lantz, B., and McKeown, N.}
\newblock {Reproducible Network Experiments using Container-based Emulation}.
\newblock In {\em CoNEXT\/} (2012).

\bibitem{longQuic}
{\sc Kakhki, A., Jero, S., Choffnes, D., Nita-Rotaru, C., and Mislove, A.}
\newblock {Taking a Long Look at QUIC: An Approach for Rigorous Evaluation of
  Rapidly Evolving Transport Protocols}.
\newblock In {\em IMC\/} (2017).

\bibitem{langley2017quic}
{\sc Langley, A., Riddoch, A., Wilk, A., Vicente, A., Krasic, C., Zhang, D.,
  Yang, F., Kouranov, F., Swett, I., Iyengar, J., et~al.}
\newblock {The QUIC Transport Protocol: Design and Internet-Scale Deployment}.
\newblock In {\em SIGCOMM\/} (2017), ACM, pp.~183--196.

\bibitem{5763591}
{\sc Millman, E., Arora, D., and Neville, S.~W.}
\newblock {STARS: A Framework for Statistically Rigorous Simulation-Based
  Network Research}.
\newblock In {\em IEEE Workshops of International Conference on Advanced
  Information Networking and Applications\/} (2011), pp.~733--739.

\bibitem{netravali2015mahimahi}
{\sc Netravali, R., Sivaraman, A., Das, S., Goyal, A., Winstein, K., Mickens,
  J., and Balakrishnan, H.}
\newblock {Mahimahi: Accurate Record-and-Replay for HTTP}.
\newblock In {\em USENIX ATC\/} (2015), pp.~417--429.

\bibitem{osterlind2006cross}
{\sc Osterlind, F., Dunkels, A., Eriksson, J., Finne, N., and Voigt, T.}
\newblock Cross-level sensor network simulation with cooja.
\newblock In {\em LCN\/} (2006), IEEE, pp.~641--648.

\bibitem{paasch2013benefits}
{\sc Paasch, C., Khalili, R., and Bonaventure, O.}
\newblock {On the Benefits of Applying Experimental Design to Improve Multipath
  TCP}.
\newblock In {\em CoNEXT\/} (2013), ACM, pp.~393--398.

\bibitem{4654311}
{\sc Perrone, L.~F., Kenna, C.~J., and Ward, B.~C.}
\newblock Enhancing the credibility of wireless network simulations with
  experiment automation.
\newblock In {\em IEEE WiMob\/} (2008), pp.~631--637.

\bibitem{Perrone:2012:SSA:2429759.2430095}
{\sc Perrone, L.~F., Main, C.~S., and Ward, B.~C.}
\newblock Safe: Simulation automation framework for experiments.
\newblock In {\em Winter Simulation Conference (WSC)\/} (2012).

\bibitem{riley2010ns}
{\sc Riley, G.~F., and Henderson, T.~R.}
\newblock {The ns-3 Network Simulator}.
\newblock {\em Modeling and tools for network simulation\/} (2010), 15--34.

\bibitem{schermann2016bifrost}
{\sc Schermann, G., Sch{\"o}ni, D., Leitner, P., and Gall, H.~C.}
\newblock {Bifrost: Supporting Continuous Deployment with Automated Enactment
  of Multi-Phase Live Testing Strategies}.
\newblock In {\em ACM/IFIP/USNIX Middleware\/} (2016), p.~12.

\bibitem{DASH2011}
{\sc Sodagar, I.}
\newblock The {MPEG-DASH} {S}tandard for {M}ultimedia {S}treaming {O}ver the
  {I}nternet.
\newblock {\em IEEE MultiMedia\/} (2011), 62--67.

\bibitem{dtarl}
{\sc Stein, M., Fr{\"o}mmgen, A., Kluge, R., Lin, W., Wilberg, A., Koldehofe,
  B., and M{\"u}hlh{\"a}user, M.}
\newblock {Scaling Topology Pattern Matching: A Distributed Approach}.
\newblock In {\em SAC\/} (2018), ACM.

\bibitem{5999877}
{\sc Stingl, D., Gross, C., Ruckert, J., Nobach, L., Kovacevic, A., and
  Steinmetz, R.}
\newblock {PeerfactSim.KOM: A simulation framework for Peer-to-Peer systems}.
\newblock In {\em IEEE HPCS\/} (2011), pp.~577--584.

\bibitem{stohr2016qoe}
{\sc Stohr, D., Fr{\"o}mmgen, A., Fornoff, J., Zink, M., Buchmann, A., and
  Effelsberg, W.}
\newblock Qoe analysis of dash cross-layer dependencies by extensive network
  emulation.
\newblock In {\em Workshop on QoE-based Analysis and Management of Data
  Communication Networks\/} (2016), ACM, pp.~25--30.

\bibitem{multimedia}
{\sc Stohr, D., Fr{\"{o}}mmgen, A., Rizk, A., Zink, M., Steinmetz, R., and
  Effelsberg, W.}
\newblock {Where are the Sweet Spots?: {A} Systematic Approach to Reproducible
  {DASH} Player Comparisons}.
\newblock In {\em ACM Multimedia\/} (2017), pp.~1113--1121.

\bibitem{varga2008overview}
{\sc Varga, A., and Hornig, R.}
\newblock {An Overview of the OMNeT++ Simulation Environment}.
\newblock In {\em Simulation tools and techniques for communications, networks
  and systems \& workshops\/} (2008), p.~60.

\bibitem{quic_submitted}
{\sc Viernickel, T., Fr\"{o}mmgen, A., Rizk, A., Koldehofe, B., and Steimetz,
  R.}
\newblock {Multipath QUIC: A Deployable Multipath Transport Protocol}.
\newblock In {\em ICC\/} (2018).

\bibitem{Zabrovskiy:2017:AAV:3083187.3083221}
{\sc Zabrovskiy, A., Kuzmin, E., Petrov, E., Timmerer, C., and Mueller, C.}
\newblock {AdViSE: Adaptive Video Streaming Evaluation Framework for the
  Automated Testing of Media Players}.
\newblock In {\em MMSys\/} (2017), ACM, pp.~217--220.

\bibitem{zeng1998glomosim}
{\sc Zeng, X., Bagrodia, R., and Gerla, M.}
\newblock {GloMoSim: a library for parallel simulation of large-scale wireless
  networks}.
\newblock In {\em Workshop on Parallel and Distributed Simulation\/} (1998),
  IEEE, pp.~154--161.

\end{thebibliography}

\end{document}